\documentclass{ws-procs9x6}
\usepackage{cite}
\newcommand{\SM}{ Standard Model}
\newcommand\openone{\leavevmode\hbox{\small1\normalsize\kern-.33em1}}

\newcommand{\tr}{ {\rm Tr} }

\newcommand{\tev}{ {\rm TeV} }

\newcommand{\lsim}{\,\raise.3ex\hbox{$<$\kern-.75em\lower1ex\hbox{$\sim$}}\,}
\newcommand{\gsim}{\,\raise.3ex\hbox{$>$\kern-.75em\lower1ex\hbox{$\sim$}}\,}

\newcommand\beq{\begin{eqnarray}}
\newcommand\eeq{\end{eqnarray}}

\begin{document}

\title{Dynamical Electroweak Superconductivity from a Composite Little Higgs}

\author{A. E. Nelson\footnote{email: anelson@phys.washington.edu}}

\address{Department of
  Physics,\\ Box 1560, University of Washington,\\ Seattle, WA 98195-1560, USA}


\maketitle

\abstracts{I describe, from the bottom up, a sequence of natural effective field theories. Below a TeV we have the minimal standard model with a light Higgs, and an extra neutral scalar. In the 1-10 TeV region  these scalars are part of a multiplet of pseudo Nambu-Goldstone Bosons (NGBs). Interactions with additional TeV mass scalars, gauge bosons, and vector-like charge 2/3 quarks stabilize the Higgs mass squared parameter without finetuning.  Electroweak superconductivity may be determined in this effective theory as a UV insensitive vacuum alignment problem.   Above the 10 TeV scale we have  strongly coupled new gauge interactions. }

\section{Introduction }
\label{intro}
A new mechanism for electroweak superconductivity, dubbed the ``little Higgs'' \cite{Arkani-Hamed:2001nc}, was recently discovered via dimensional deconstruction \cite{Arkani-Hamed:2001ca,Hill:2000mu}. This mechanism  has since been realized in various simple nonlinear sigma models \cite{Arkani-Hamed:2002qy,Arkani-Hamed:2002qx,Low:2002ws,Wacker:2002ar,Kaplan:2003uc,Chang:2003un}. 
In this talk, I review some of these developments, and describe some work in progress on incorporating the little Higgs mechanism in a strongly coupled model of dynamical symmetry breaking, in which the Higgs, part of the top and part of the left handed bottom are composite particles\cite{us}.

According to an old proposal of Georgi and Pais \cite{Georgi:1975tz}, the Higgs is an approximate Nambu-Goldstone Boson (NGB), whose mass squared is protected against large radiative corrections by approximate nonlinearly realized global symmetries.   Of course, saying that the Higgs is a pseudo NGB is not enough to explain a small mass squared,  because  the Yukawa, self, and gauge interactions explicitly break any nonlinearly realized symmetry and lead to quadratic sensitivity of the Higgs mass squared to short distance physics.   However little Higgs theories realize the NGB proposal  in a UV insensitive way.  In these theories the Yukawa, gauge and self couplings arise due to the combined efforts of a collection of symmetry breaking terms in the effective theory at 1 TeV. The Higgs mass squared is at most logarithmically sensitive to the cutoff at one loop, provided the symmetry breaking terms  satisfy a mild criterion: {\it no {\bf single}  term in the Lagrangian breaks {\bf all} the  symmetry which is protecting the Higgs mass}.  Such symmetry breaking may be thought of as being ``nonlocal in theory space'' and is softer than usual. Several new, {\bf weakly coupled} particles are found around a few TeV and below, which cancel the leading quadratic divergences in the Higgs mass in a manner reminiscent of softly broken supersymmetry. However, unlike supersymmetry, the cancellations occur between particles of the same statistics, and there is a natural ``little  hierarchy'', of order $\lambda/4\pi$, between the $W$ mass scale and the scale of the new physics. The spectrum and phenomenology of little Higgs theories has been discussed in refs. \cite{Arkani-Hamed:2002pa,Gregoire:2002ra,Han:2003wu,Burdman:2002ns}.  

In some little Higgs theories,   corrections to precision electroweak observables  are comparable in size to one-loop standard model effects over much of the natural parameter space\cite{Csaki:2002qg,Hewett:2002px,Chivukula:2002ww,Han:2003wu}, providing important constraints. It is, however,  straightforward to find natural, simple,  and experimentally viable little Higgs theories where the corrections are much smaller\cite{Chang:2003un,Csaki:2003si}. This will be explicitly discussed in ref. \cite{us}.

A pressing issue   is to situate the little Higgs in a more complete theory with a higher cutoff.  This is necessary  to address in a compelling way the  phenomenology of flavor changing neutral currents\cite{Lane:2002pe,Chivukula:2002ww}, which is sensitive to physics beyond  10 TeV.
Here I describe how to embed a slightly altered  version of the ``littlest'' \cite{Arkani-Hamed:2002qy} Higgs model into a UV complete theory.  The model is experimentally viable, with acceptable precision electroweak corrections and no more than about 10\%
 fine tuning.  The Higgs is a composite of fermions interacting via strong dynamics at the 10 TeV scale. The correct size of quark and lepton masses can be generated  without excessive flavor changing neutral currents from four fermion couplings. Generating these interactions require either introducing heavy particles of mass between 30 and 250 TeV (depending on how strongly coupled they are) or  one might conceivably postpone the issue of generation of 4 fermion  coupling further if the theory possesses significant anomalous scaling in the UV. 
 Note that many  UV completions of little Higgs theories are conceivable, and the  model discussed here should by no means be taken as canonical.

This model illustrates several advantages of composite little Higgs models as compared with the traditional approaches to electroweak superconductivity. In contrast to the Minimal Supersymmetric Standard Model (MSSM), there are no problems with fine tuning, lepton flavor violation, CP violation, or flavor changing neutral currents. The natural expectation for the masses of the visible new particles is well above the weak scale. The chief drawback of the model  relative to the MSSM is the lack of a prediction for the weak angle. In contrast to Technicolor, there is a light Higgs in the the low energy spectrum and it is straightforward to make the precision electroweak corrections very similar to those in the Standard Model.  In contrast to the minimal Standard Model, the Higgs mass squared parameter is not UV sensitive, does not require fine tuning, and the dynamics driving the Higgs condensate occurs at and below 10 TeV, where it can be studied experimentally.
In contrast  to the Georgi-Kaplan composite
Higgs\cite{Kaplan:1984fs,Kaplan:1984sm,Georgi:1984af,Georgi:1984ef,Dugan:1985hq},  the hierarchy between the compositeness scale and the electroweak scale is
not due to  fine tuning of  parameters.  

I will describe this model from the bottom up, as a sequence of natural effective field theories. Below a TeV, the effective theory is the minimal standard model, with an additional light neutral scalar and higher dimension operators with small coefficients. 

\section{The $SU(5)/SO(5)$  ``littlest'' Higgs Model}
\label{littlest}

At the TeV scale,  we embed this theory into a nonlinear sigma model.
Here I will discuss
the simplest of the little Higgs models, the ``littlest Higgs''.  The target space is  the coset space $SU(5)/SO(5)$ \cite{Arkani-Hamed:2002qy}. 
This theory describes the low energy interactions of 14 NGBs, with decay constant $f\sim1$~ TeV. At 1 TeV, all interactions in the effective theory are weak.    The cutoff of this effective theory is about $4\pi f\sim$10~TeV, where the  NGB interactions become strong. The SU(5) symmetry is explicitly broken by  gauge interactions and fermion couplings,  leading to masses for most of the NGBs of order $f$, while others get eaten by  gauge bosons whose mass is also of order $f$. A special subset of the NGBs, however, do not receive masses to leading order in the symmetry breaking terms, and are about a factor of $4\pi$ lighter than $f$. In the minimal model of ref. \cite{Arkani-Hamed:2002qy}, this subset consisted only of a single Higgs doublet,  dubbed the little Higgs. In the present model a neutral scalar also remains   light. At the TeV scale a small
number of additional scalars, vector bosons and quarks cancel the one
loop quadratic divergence in the Higgs mass without fine tuning or
supersymmetry.  

We can describe the $SU(5)\rightarrow SO(5)$ breaking as arising from a
vacuum expectation value for a $5\times 5$ symmetric matrix $\Phi$,
which transforms as $\Phi \to V \Phi V^T$ under $SU(5)$.  As we will see later,  this $\Phi$ corresponds in the theory above 10 TeV to a fermion bilinear of a strongly coupled gauge theory. A vacuum
expectation value for $\Phi$ proportional to the unit matrix 
breaks $SU(5) \to SO(5)$. For later convenience, we use an equivalent
basis where the vacuum expectation value for the symmetric tensor
points in the $\Sigma_0$ direction where $\Sigma_0$ is
\begin{equation}
  \label{sigma}
  \Sigma_0=\left(
  \begin{array}{ccc}
    \quad  & \quad & \openone \\ \quad & 1 &\quad \\ \openone
    &\quad&\quad 
  \end{array}\right)
  \ .
\end{equation}
The unbroken $SO(5)$ generators satisfy 
\begin{equation}
  T_a \Sigma_0 + \Sigma_0 T_a^T = 0
\end{equation}
while the broken generators obey
\begin{equation}
  \label{brokengen}
  X_a \Sigma_0 - \Sigma_0 X_a^T = 0\ .
\end{equation}
The Goldstone bosons are fluctuations about this background
in the broken directions $\Pi \equiv \pi^a X^a$, and can be
parameterized by the non-linear sigma model field
\begin{equation}
  \Sigma(x) = e^{i \Pi/f} \Sigma_0 e^{i \Pi^T/f} = 
  e^{2 i \Pi/f}  \Sigma_0,
\end{equation}
where the last step follows from eq. \ref{brokengen}.

We now introduce the gauge and Yukawa interactions which explicitly
break the global symmetry. In ref. \cite{Arkani-Hamed:2002qy}, these
were chosen to ensure an  $SU(3)$ global symmetry under which the little Higgs transformed nonlinearly, in the limit
where any of the couplings were turned off.   This required embedding the electroweak  $SU(2)_w\times U(1)_y$ gauge interaction  into an $[SU(2)\otimes U(1)]^2$ gauge group, which was spontaneously broken to $SU(2)_w\otimes U(1)_y$ at the scale $f$. This led to a rather light $Z'$,  which was constrained by Tevatron and  precision electroweak corrections \cite{Csaki:2002qg,Hewett:2002px,Han:2003wu}. 
Due to the small size of the weak angle, we can eliminate the $Z'$ and the associated constraints without increasing the finetuning of the theory. With a 10 TeV cutoff, naturalness does not require  cancellation of its quadratically cutoff sensitive contribution to the Higgs mass squared from weak hypercharge gauge  interactions.  We therefore only introduce a single U(1). This simplification will make cancellation of gauge anomalies very simple in the underlying composite model, as well as relaxing experimental constraints. 
We thus weakly gauge an $  SU(2)^2 \times U(1)_y$ subgroup of the $SU(5)$ global
symmetry. The generators of the $SU(2)$'s are embedded into $SU(5)$ as
\begin{equation}
  Q_1^a =
  \pmatrix{
    \sigma^a/2 &\quad&\quad\cr  \quad&\quad&\quad\cr
    \quad&\quad&\quad
  }, \\
 Q_2^a =
  \pmatrix{ \quad&\quad&\quad\cr
    \quad&\quad&\quad\cr
     \quad&\quad&-{\sigma^a}^*/2
 }, \end{equation}
 while the generators of the $ U(1)$ are
    given by\begin{equation}
  Y = {\rm diag}(1,1,0,-1,-1)/2 \ .
\end{equation}
In this basis the 14 NGBs have definite electroweak quantum numbers. We write the NGB matrix as
\begin{equation}
  \label{pgb} 
  \Pi=
  \pmatrix{
   0 &\frac{h^\dagger}{\sqrt{2}}&\phi^\dagger\cr
    \frac{h}{\sqrt{2}}&0
    &\frac{h^*}{\sqrt{2}}\cr
    \phi&\frac{h^T}{\sqrt{2}}&    0
 }\end{equation}
where $h$ is the Higgs doublet, $h=(h^+, h^0)$, and $\phi$ is an electroweak triplet carrying one unit of weak hypercharge, represented
 as a symmetric two by two matrix.  In eq. \ref{pgb}, I have ignored the three
Goldstone bosons that are eaten in the Higgsing of $SU(2)^2\times
U(1) \to SU(2)\times U(1)$, as well as an additional neutral NGB which is massless. (In order to avoid phenomenological problems from a massless NGB, which, for instance, is constrained by rare kaon decays,  we can add small symmetry breaking terms to the potential in order to give the NGB a small mass, without  affecting the discussion of the little Higgs).

The effective theory at the scale $f$ has a tripartite
 tree-level Lagrangian,  given by 
\begin{equation}
  \label{eq:lag}
  { L} = { L}_K + { L}_t + { L}_\psi .
\end{equation}
Here ${ L}_K$ contains the kinetic terms for all the fields;
${ L}_t$ generates the top Yukawa coupling; and ${ L}_\psi$
generates the remaining small Yukawa couplings.  I describe each  of these pieces in turn.

The kinetic terms include  the usual kinetic terms for gauge and Fermi fields. The leading two-derivative term for the
non-linear sigma model is 
\begin{equation}
L_K\supset  \frac{f^2}{8} \tr{D_\mu\Sigma}D^\mu \Sigma^\dagger
\end{equation}
where the covariant derivative of $\Sigma$ is given by 
\begin{equation}
  \label{eq:cd}
  D\Sigma = \partial \Sigma - \sum_j \left\{
   i g_{j} W_j^a (Q_j^a \Sigma + \Sigma Q_j^{aT}) 
  + i g'_ B(Y \Sigma + \Sigma Y^T)\right\}.
\end{equation}
The $g_i, g^\prime$ are the couplings of the $SU(2)^2\times
  U(1)$ groups.  This term will result in the leading quartic term in the Higgs potential, as I will describe shortly.

Generation of the top Yukawa coupling while preserving UV insensitivity requires  new heavy fermions,  in addition to the usual third-family weak
doublet  quarks $q_3 = (t,b')$ and weak singlet $\bar u_3$.
  In ref. \cite{Arkani-Hamed:2002qy} it was shown that a large top Yukawa
coupling could be included without inducing a large quadratic divergence in
the Higgs mass by simply  adding a pair of colored left handed Weyl Fermions
$\tilde{t},\tilde{t}^c$,  transforming as a singlet under weak $SU(2)$.
That choice was minimal in new particle content. Here I will describe a different choice, which is more natural  in composite model building, where we expect the $SU(5)$ symmetry to arise as an accidental symmetry of the  dynamics of a strongly coupled theory, analogous to the
$SU(3)\times SU(3)$ chiral symmetry of QCD. Composite fermions  will naturally couple to the composite bosons.
I therefore  introduce new ``composite'' fermions $X,\bar X$ transforming as $(5,3)$ and $(5, \bar 3)$, respectively, under
$SU(5)\times SU(3)_{c}$.  
 These couple to the $\Sigma$ field
in an $SU(5)$ symmetric fashion and gain mass from the $SU(5)/SO(5)$ symmetry breaking. The top mass  will arise by mixing ``fundamental'' quarks  $q_3$ and $\bar u_3$ with ``composite'' quarks of the same quantum numbers, in a manner  similar to Frogatt-Nielsen models of flavor
\cite{Froggatt:1979nt} and the top see-saw \cite{Dobrescu:1998nm,
  Chivukula:1998wd}.
  
  Explicitly,  the fields $X,\bar X$, contain components $\tilde q, \tilde t,p,\bar p,\bar{\tilde t}, \bar {\tilde q}$, transforming under $SU(3)_c\times SU(2)_1\times SU(2)_2\times U(1)_y$ as

 \begin{tabular}{|r|r|cccc|}
\hline
  & & $SU(3)_c$ & $SU(2)_1$&$SU(2)_2$&$U(1)_Y $\\
\hline
& $\tilde q$& 3&2&1&1/6\\
X&$\tilde t$& 3 & 1 & 1 & 2/3\\
&$p$  &3&1 & 2 &  7/6 \\
\hline
&$\bar p$  &$\bar3$&2 &1  &  -7/6 \\
$\bar X$&$\bar{\tilde t}$& $\bar3$ & 1 & 1 & -2/3\\
& $\bar {\tilde q}$& $\bar3$&1&2&-1/6\\
\hline
\end{tabular}

We break the $SU(5)$ symmetry only through explicit fermion
mass terms connecting the $q_3$ and $\bar u_3$ to the components of $X, \bar X$ with the appropriate quantum numbers.
The top Yukawa coupling arises from  the combination of terms

\begin{equation}
  \label{topmass1}
  L_{t}=\lambda_1 \bar X \Sigma^\dagger X + \lambda_2 f \bar{\tilde q} q_3 + \lambda_3  f \bar u_3 \tilde  t + {\rm h.c.}
  \end{equation}

The approximate global symmetry of this effective Lagrangian is actually $SU(5)^3$, with independent $SU(5)$'s acting on $\Sigma$, $X$, and $\bar X$.
The first term breaks the three $SU(5)'s$ to the diagonal subgroup, while the second and third terms each leave  two of the three   $SU(5)$'s unbroken.
Because all three terms are needed to entirely break  the symmetry protecting the little Higgs mass, this form of symmetry
breaking is soft enough to not induce quadratic or logrithmic divergences at one
loop, or quadratic divergences at two loops.

To see that ${ L}_t$
generates a top Yukawa coupling we expand ${ L}_t$ to first order
in the Higgs $h$:
\begin{equation}
  { L}_t \supset \lambda_1  \bar{\tilde t}q_3 h + f(\lambda_1  \bar{\tilde t} + \lambda_3 \bar u_3)\tilde{t}  +f \bar{\tilde q}( \lambda_1 \tilde q+ \lambda_2 q_3)
+ \cdots\ .
\end{equation}
Clearly $\tilde{t}$ marries the linear combination $(\lambda_1  \bar{\tilde t} + \lambda_3 \bar u_3)/(\lambda_1^2+\lambda_3^2)^{1/2}$  to become massive,  $\bar{\tilde q}$ marries  the   linear combination $( \lambda_1 \tilde q+ \lambda_2 q_3)/(\lambda_1^2+\lambda_2^2)^{1/2}$, and $p$ pairs up with $\bar p$.  We can integrate out these heavy quarks.
The remaining light combinations are $Q$, the  left handed top and bottom doublet, \begin{equation}
Q \equiv\frac{(\lambda_2 \tilde q- \lambda_1 q_3)}{\sqrt{\lambda_1^2+\lambda_2^2}}\ ,
\end{equation}
and $\bar T$, the left handed antitop,
\begin{equation}
\bar T \equiv \frac{(\lambda_3 \bar{ \tilde t}- \lambda_1 \bar {u}_3)}{\sqrt{\lambda_1^2+\lambda_3^2}}\ ,
\end{equation}
with a Yukawa coupling to the little Higgs
\begin{equation}
  \label{topyuk}
  \lambda_t \, h\bar U Q +{\rm h.c.} \qquad {\rm where} \qquad
  \lambda_t = \frac{\lambda_1 
    \lambda_2\lambda_3}{\sqrt{\lambda_1^2 + \lambda_2^2}\sqrt{\lambda_1^2 + \lambda_3^2}}\ .
\end{equation}

Finally, the interactions in ${ L}_\psi$ encode the remaining
Yukawa couplings of the \SM. 
\begin{equation}{ L}_\psi=\lambda^e_{ij} \bar e_i \ell_j h+\lambda^d_{ij} \bar d_i q_j h + \lambda^u_{ij} \bar u_i q_j h^\dagger + {\rm h.c.,}\end{equation} where in the third term all coupling s are very small and not the major source of the top Yukawa coupling to the Higgs.  These couplings are explicitly SU(5) breaking but small enough so that the
1-loop quadratically divergent contributions to the Higgs mass they
induce are negligible with a cutoff $\Lambda_\chi \sim {10\ \rm TeV}$.  

 Note that since
there may be additional fermions at the cutoff which cancel the
anomalies involving the broken subgroup  we need insist only that \SM\ 
anomalies cancel in the effective theory at the TeV scale. It is however, simple to write down an entirely anomaly free theory at the 10 TeV scale.

We now turn to a discussion of loop effects in this effective theory, which give the  Higgs a potential.

\subsection{The Effective Potential and Electroweak Superconductivity}
At tree level the orientation of the $\Sigma$ field is undetermined, and all the NGBs are massless. Whether or not we have electroweak superconductivity is  a problem of vacuum alignment, which can be settled by a computation of the Higgs effective potential at one loop order.
Our nonrenormalizable effective theory is incomplete, and we will need to add new interactions (counterterms) in order to account for the cutoff sensitivity introduced by radiative corrections.   We follow a standard chiral Lagrangian analysis,  including all
operators consistent with the symmetries of the theory with 
coefficients  assumed to be of the order determined by na\"\i{}ve dimensional
analysis\cite{Manohar:1984md,Cohen:1997rt,Luty:1998fk}, that is, of similar size to the radiative corrections computed from the lowest order terms with cutoff $\Lambda_\chi=4\pi f$.
Remarkably, the leading such terms only contribute to the quartic term in the Higgs potential, and not the quadratic term.

The largest corrections  come from the gauge sector, due to  1-loop quadratic divergences proportional to
\begin{equation}  \frac{\Lambda_\chi^2}{16 \pi^2} \tr M_V^2(\Sigma)
\end{equation}
where $M^2(\Sigma)$ is the gauge boson mass matrix in a background
$\Sigma$. $M_V^2(\Sigma)$ can be read off from the covariant
derivative for $\Sigma$  of eq. \ref{eq:cd}, giving a potential
\begin{equation}
  \label{eq:cw1}
  c g^2_{j} f^4 \sum_a \tr \left[(Q_j^a \Sigma)(Q_j^a \Sigma)^*\right]
  + c {g'}^2 f^4 \tr \left[(Y \Sigma)(Y \Sigma)^*\right]
\end{equation}
Here  $c$ is an ${ O}(1)$
constant whose precise value is sensitive to the UV physics at the
scale $\Lambda$.  Note that at second
order in the gauge couplings and momenta eq. \ref{eq:cw1} is the unique
gauge invariant term transforming properly under the global $SU(5)$
symmetry.  This potential is analogous to that generated by
electromagnetic interactions in the pion chiral Lagrangian, which
shift the masses of $\pi^\pm$ from that of the $\pi^0$
\cite{Das:1967it}. In analogy to the chiral
Lagrangian, we assume that $c$ is positive. This implies that the gauge interactions prefer the alignment $\Sigma_0$ where the electroweak group remains unbroken. 

In the following, for simplicity,  we neglect effects which are suppressed by the weak angle sin$^2\theta_W$.
To quadratic order in $\phi$ and quartic order in $h$, the potential from the SU(2) gauge interactions of  eq. \ref{eq:cw1} is
\begin{equation}
  \label{eq:hpot}
  + c  g_1^2f^2 |{\phi_{ij} - \frac{i}{2f} (h_i h_j +
    h_j h_i)}|^2 +
    c g_2^2 f^2 |{\phi_{ij} + \frac{i}{2f} (h_i h_j  + h_j h_i)}|^2 )\  .
\end{equation}
The SU(2) interactions in eq.  \ref{eq:cw1} gives the triplet a positive mass squared of order
\begin{equation}
m_\phi^2= c(g_1^2+g_2^2 ) f^2\ .
\end{equation}
 The little Higgs doublet, however, only receives mass at this order form the $U(1)_Y$ interactions,  because the $SU(2)_{1,2}$ gauge interactions  each leave  an $SU(3)$
symmetry intact, under which the little Higgs transforms nonlinearly \cite{Arkani-Hamed:2002qy}.
The SU(2) interactions do, however, lead to an effective quartic interaction term in the little Higgs potential, as well as interaction with the $\phi$ triplet.
After the Higgs triplet is integrated out, the resulting quartic coupling for the little Higgs from the SU(2) interactions is
\begin{equation}
\lambda=c\frac{g_1^2g_2^2}{(g_1^2+g_2^2)} .
\end{equation}
Remarkably,  the SU(2) interactions do not lead to a mass squared for the little Higgs at this order, although they do give a quartic term in the Higgs potential which is of order 1.  

The remaining part of the vector boson contribution to the
Coleman-Weinberg potential is
\begin{equation}
\label{eq:cw2}
  \frac{3}{64\pi^2}\tr
    M_V^4(\Sigma)\log
      \frac{M_V^2(\Sigma)}{\Lambda_\chi^2}\ .
\end{equation} 
This gives a logarithmically enhanced positive Higgs mass squared from the $SU(2)$ interactions 
\begin{equation} 
\label{eq:vector}
  \delta m_h^2=\frac{9 g^2 {M'_W}^2}{64\pi^2} 
 \log\frac{\Lambda_\chi^2}{{M'_W}^2}
\end{equation} 
where $M'_W$ is the mass of the heavy $SU(2)$ triplet of gauge bosons. 
There is a similar Coleman-Weinberg potential from the scalar
self-interactions in eq. \ref{eq:cw1} which also
give logarithmically enhanced positive contributions to the Higgs mass
squared:
\begin{equation}
  \label{eq:2}
 \delta m_h^2= \frac{\lambda}{16\pi^2}M_\phi^2
  \log\frac{\Lambda_\chi^2}{M_\phi^2}
\end{equation}
where $M_\phi$ is the triplet scalar mass.

In this theory, as in the MSSM, the top drives 
electroweak symmetry breaking.  
A negative mass squared term in the Higgs potential 
 comes from  the fermion  loop contribution to the
Coleman-Weinberg potential, which  is
\begin{equation}
\label{topcw}
     - \frac{3}{16\pi^2} \tr \left(M_f(\Sigma)
      M_f^\dagger(\Sigma)\right)^2\log\frac{M_f(\Sigma)
          M_f^\dagger(\Sigma)}{\Lambda_\chi^2}
\end{equation}
where $M_f(\Sigma)$ is the fermion mass matrix in a background
$\Sigma$.  We can neglect the contributions of the light fermions to this potential, and only consider the effects of the heavy charge 2/3 quarks contained in $\tilde{t},\bar{\tilde{t}},\tilde{q_t},\bar{\tilde{q_t}},p_t,\bar{p}_t,q_t$ and $\bar{u_3}$. Here $q_t, \tilde{q_t},\bar{\tilde{q_t}},p_t,\bar{p}_t$ denote the charge 2/3 components of the respective weak doublets. 

The charge 2/3 quark mass matrix is

 \begin{tabular}{|r|cccc|}
\hline
M   & $p_t$ & $\tilde{t}$&$\tilde{q}_t$&$q_t $\\
\hline
$\bar{p}_t$& $\lambda_1 f\cos^2\theta$& $\lambda_1 f \frac{i }{\sqrt2}\sin2\theta$&$-\lambda_1 f\sin^2\theta$&0\\
$\bar{\tilde{t}}$&$\lambda_1 f \frac{i }{\sqrt2}\sin2\theta$& $\lambda_1 f \cos 2\theta$& $\lambda_1 f \frac{i }{\sqrt2}\sin2\theta$ & 0\\
$\bar{\tilde{q_t}}$&$-\lambda_1 f\sin^2\theta$  &$\lambda_1 f \frac{i }{\sqrt2}\sin2\theta$&$\lambda_1 f\cos^2\theta$ & $\lambda_3 f$ \\
$\bar{u_3}$&0  &$\lambda_2 f$&0&0 \\
\hline
\end{tabular}

\noindent where $\theta=\langle h\rangle/(\sqrt2 f)$.
Note that
\begin{equation}
\frac{\partial}{\partial \theta}{\rm Tr} M^\dagger M=0\end{equation}
and \begin{equation}\frac{\partial}{\partial \theta}{\rm Tr}  (M^\dagger M)^2=0\end{equation}
which guarantees cutoff insensitivity of the one loop radiative corrections to the little Higgs potential from this sector.
Besides the top which has mass $\lambda_t \langle h\rangle$, there are three heavy quarks, of mass 
\beq
M_1&=&\lambda_1 f \cr 
M_2&=&\left(a^2 +\frac{ \lambda_t^2\langle h\rangle^2b^2}{a^2 - b^2}  +O(\langle h\rangle ^4)\right)^{1/2}\cr
M_3&=&\left(b^2  -\frac{\lambda_t^2\langle h\rangle^2a^2 }{a^2 - b^2}  +O(\langle h \rangle ^4)\right)^{1/2}
\eeq
where
\beq
a^2&=&(\lambda_1^2+\lambda_2^2) f^2 \cr
b^2&=&(\lambda_1^2+\lambda_3^2) f^2 \ , \eeq and we have assumed $\lambda_2\ne \lambda_3$ so that nondegenerate perturbation theory is appropriate for diagonalizing the quark masses.
We denote these three heavy charge 2/3  quarks as the the $t',t'', t'''$,  respectively. Note that if mixing terms of order $h/f$ are neglected, these quarks have vector-like standard model gauge quantum numbers (3,2,7/6), (3,1,2/3), and (3,2,1/6) respectively.
Including the top and   $t'', t'''$ in equation \ref{topcw} gives a cutoff insensitive contribution to the little Higgs effective potential
\goodbreak
\beq
  \label{topx}\delta_{V_{\rm eff} }=
    &&-\frac{3\lambda_t^2 h^2}{8 \pi^2}
     \frac{a^2b^2}{a^2-b^2}\log\left(\frac{a^2}{b^2} \right)\cr
     &&+ \frac{3\lambda_t^4 h^4}{16 \pi^2} \Bigg( \frac{(a^2+b^2)\left((3a^4+3b^4-4b^2a^2)\log\left(\frac{a^2}{b^2}\right)- (a^4-b^4)\right)} {2(a^2-b^2)^3}\cr &&+ \log\left( \frac{ab}{h^2}\right) \Bigg)+ O(h^6)\  . 
  \eeq
Note that the contribution to the mass squared is negative, and typically of somewhat larger magnitude than the positive gauge contribution. 
For  $\lambda_1\sim \lambda_2\sim \lambda_3\sim 2$ we have a top Yukawa coupling of order 1, and a reasonably sized contribution to the quadratic term in the potential from the top sector.
The top sector contribution to the quartic term is  positive, and logarithmically enhanced. Although this  is parametrically of higher order than the quartic term  from the gauge sector, numerically it is comparable. It is straightforward to find values of $a$, $b$, $c$, and $f$ which give the correct Higgs vev without significant fine tuning. Parametrically, $f\sim 4\pi M_W/(\sqrt{N_c}\lambda_t^2)$, and the masses of new heavy particles should  naturally be of order a few TeV.

\section{The little Higgs as a Composite Higgs}
\label{comp}

We now turn to  the effective theory above  10 TeV. We assume The $SU(5)/SO(5)$ symmetry breaking pattern arises from condensation of a new set of   fermions, called
Ultrafermions, which transform in a real representation of a new strong gauge group, called Ultracolor \cite{Dugan:1985hq}. This will result in composite  NGBs,  like the pions of QCD.
 For concreteness, we take Ultracolor to be an $SO(7)$ gauge group. 
  
\subsection{Matter content }
We thus assume that above 10 TeV we have an $SO(7)\times SU(3)\times SU(2)'\times SU(2)\times U(1) $
gauge theory, with  the fermion matter content of the following table:
\newpage
\begin{center}
{ {\bf  Fermions }}
\end{center}
\begin{tabular}{|r|c c c  cc|}
\hline
  & $SO(7)$& $SU(3)_c$ & $SU(2)' $&$SU(2)$&$U(1)_Y $\\
\hline
$\bar e_i$ & 1 & 1 & 1 &1& 1 \\
$\ell_i$ & 1 & 1 & 1 &2& -1/2 \\
$q_i$& 1 & 3 & 1&2 & 1/6\\
$\bar u_i$  &1&$\bar 3$ & 1 & 1 & -2/3 \\
$\bar d_i$&1& $\bar 3$ & 1 & 1 & 1/3 \\ 
$\lambda$ & 21 & $ 1$ & 1 &1& 0\\ 
$\phi_{\bar 3}$ & 7 & $ \bar 3$ & 1 & 1&-2/3 \\ 
$\phi_{ 3}$ & 7 & $ 3$ & 1 & 1&2/3 \\ 
$\phi_{2'}$&7&1&2&1&-1/2\\
$\phi_{ 2}$&7&1&1&2&1/2\\
$\phi_{ 0}$&7&1&1&1&0\\
$\bar\psi$&1&$\bar3$&1&2&-7/6\\
$\psi$&1&3&2&1&7/6\\
\hline
\end{tabular}\

Here $i=1,2,3$ is a generational index. The conjectured dynamics of the strong $SO(7)$ gauge interaction will be discussed below. 
The only role of the fields $\psi$,  and $ \bar\psi$ is to cancel  $SU(2)^2 U(1)$ and $SU(2)'^2 U(1)$ anomalies. Note that this theory is free of gauge anomalies.

The approximate SU(5) global symmetry of the littlest Higgs nonlinear sigma model acts on the fermions $\phi_2$, $\phi_{2'}$, and $\phi_0$. This symmetry is explicitly broken by the $SU(2)'\times SU(2)\times U(1)_Y$ gauge interactions and by  four fermion operators.

In order to account for quark, lepton and $\psi,\bar\psi$ masses, we assume  the effective Lagrangian contains terms:
\beq \label{four} L &  \supset & m_{\lambda} \lambda\lambda+ m_3 \phi_{\bar 3} \phi_3+ m_0 \phi_0\phi_0+ h_q Q \phi_{\bar 3} \phi_2\lambda + h_{\bar u} \bar U \phi_3\phi_0\lambda+ h'_s \psi  \phi_{\bar 3}\phi_{2'}   \lambda \cr & & +
h_s \bar\psi\phi_3\phi_{2}\lambda + h^u_{ij}\phi_2\phi_0 \bar u_i q_j  +h^d_{ij} (\phi_2\phi_0)^\dagger \bar d_i q_j+h^e_{ij} (\phi_2\phi_0)^\dagger \bar e_i \ell_j\cr &&
+{\rm h.c.}
\eeq

The mass terms $m_{\lambda},\ m_3, $ and $m_0$ are small compared to the $SO(7)$ strong coupling scale. Only $m_3$   plays an important role in the dynamics of the theory. $m_0$ and $m_\lambda$ give mass to  otherwise dangerous axions.  $m_0$ can be anywhere between a  GeV and   about a  hundred GeV, while $m_\lambda$ could be as large as  a few TeV. $m_3$ is assumed to be about a TeV.  The $h_q$ and $h_{\bar u}$ terms are going to lead to the seesaw top quark mass. The four fermi coupling constants $h^u, h^d,$ and $h^e$ are small, and the light fermions are very weakly coupled to the strong dynamics.

\subsection{ Dynamical Assumptions }

Here I describe the dynamical assumptions which will lead to the low energy effective theory of the previous section.

Take the SO(7) gauge interaction to be  confining at a scale $\Lambda$ which is at or above the chiral symmetry breaking scale  $\Lambda_\chi\sim 4\pi f$. Neglecting all weak interactions and the terms in eq. \ref{four}, the global symmetries of the 
theory  are an SU(11) which acts on the 11 fermions in the  fundamental representation of SO(7) (all the $\phi$ fields) and an 
anomaly free U(1), carried by $\lambda$ as well  as the $\phi$ fields. The 'tHooft anomaly matching conditions \cite{thooft} require either spontaneous symmetry breaking or massless composite fermions. There are no simple massless fermion solutions to all the 'tHooft anomaly matching conditions for the  $SU(11)\times U(1)$, so it is  expected that  at least part of the global symmetry is spontaneously broken from fermion condensates.  We assume a $\lambda\lambda$ condensate, spontaneously breaking the U(1). It is conceivable 
that  SU(11)  is  spontaneously broken to SO(11) by a $\phi\phi$ condensate, but it is also possible to match the 
'tHooft conditions in a simple way, with massless composite  spin 1/2 fermions formed of 
$\phi\phi\lambda$, in an antisymmetric tensor representation of SU(11). Note that the anomaly of the antisymmetric tensor of the SU(11) is 7, so 
such massless bound states  match the SU(11) anomaly of the fundamental fermions. We therefore make the reasonable  assumption that the $\phi\phi$ condensate does not form, and, in the absence of the terms in eq. \ref{four}, the SU(11) is not spontaneously broken.

If there were   fewer fermions, this simple anomaly matching will not work, and we might expect confinement to trigger chiral symmetry breaking.  Furthermore the mass term $m_3$ explicitly breaks the SU(11) symmetry to $SU(5)\times SU(3)$, and  some of the composite 
fermions  which are massless in order to match the SU(5) anomalies contain $\phi_3$ and $\phi_{\bar 3}$ as constituents.  It therefore seems likely that if the mass term $m_3$  becomes too large, the remaining SU(5) chiral symmetry must  spontaneously 
break to SO(5).  We assume this happens with $m_3\sim $  a few TeV.  All the composite fermions will then acquire a mass.  
In particular,  the composites $X,\bar X$,  of the previous section which s transform as $(5,3)$ and $(5,\bar{3}))$,  are made from $\lambda\phi_3\phi_2,\  \lambda\phi_3\phi_{2'},\  \lambda\phi_3\phi_0,\ \lambda\phi_{\bar 3}\phi_2,\ \lambda\phi_{\bar 3}\phi_{2'},\ \lambda\phi_{\bar 3}\phi_0$. However, in our case 
the mass of $X$, $\bar X$, besides being proportional to the $SU(5)$ chiral symmetry breaking scale, must also contain the mass term $m_3$ which breaks the approximate
$SU(6)$ symmetry acting  on $\phi_3$ and $\phi_{\bar 3}$.  Thus, it should be proportional to 
$4\pi f \frac{m_3}{\Lambda} $. 
The coupling $\lambda_1$ of the previous section,  should be of order $4 \pi m_3/\Lambda$.
  We  therefore take  $m_3$ to be a few TeV.  
  
Using naive dimensional analysis, we find  the couplings $\lambda_{2,3}$  of the previous section are, respectively, of order $h_{q,\bar u} \Lambda^3/(16 \pi^2 f)$. A toy set of assumptions, which we discuss more fully in \cite{us}, suggests that a reasonable value for $\Lambda $ is about 50 TeV, and that  for this value of $\Lambda$ a mass $m_3$ of a few TeV can drive chiral symmetry breaking provided the spectrum of  spinless mesons below 50 TeV contains a scalar in the symmetric tensor representation of SU(11) of mass $\sim$10 TeV.  Such an apparently unnaturally light scalar   could result from  the number of flavors of the SO(7) theory being  very near the critical number of flavors which divides a confining from a conformal phase. For $\Lambda =10, 50$ TeV,  the couplings $\lambda_{2,3}$ will be of order 2 when the coefficient of the four fermi coupling is 
$h_{q,\bar u} \sim (4\pi)^2/(30\ \tev)^2  ,  (4\pi)^2/(250\ \tev)^2$ respectively. Such four fermi couplings will either require additional, strongly coupled fields of mass $\lsim 30,250$~TeV, or substantial anomalous scaling above $\Lambda$. Flavor changing neutral currents are not   a  problem provided either these new fields couple weakly to the light quarks and leptons, or provided anomalous scaling   or a high compositeness scale allows the new fields to be sufficiently heavy. A more thorough discussion of these issues will be presented in ref. \cite{us}.
\section{Recap }
\label{conclusions}
We have presented a sequence of natural effective field theories, with no severe finetuning or phenomenological difficulties, describing electroweak symmetry breaking. The underlying theory is a  strongly coupled, perhaps nearly conformal theory, valid to  some very high energy scale.  At some scale above 10 TeV, perhaps of order 50 TeV, a  mass term for some fields steers the theory into a confining phase, with an unbroken approximate SU(11) chiral symmetry and relatively light composite fermions.  At 1 TeV,  another mass term explicitly breaks  the   SU(11) chiral symmetry  down to SU(5), and drives spontaneous breaking of   SU(5) symmetry to  SO(5). All the light composite fermions obtain mass at this scale. Most of the resulting pseudo-Goldstone bosons get mass from explicit symmetry breaking at the TeV scale, or are eaten by TeV mass gauge bosons. The exception is the little Higgs, a doublet which receives a small, ultraviolet-insensitive negative mass squared from  loops in the top quark mass sector.  Although this Higgs is a composite particle, it acts like a weakly coupled elementary scalar in the effective theory, whose condensate produces electroweak superconductivity.

This theory provides an example of dynamical symmetry breaking, which phenomenologically resembles the minimal Standard Model at low energies. At the TeV scale, it distinguishes itself via   a  new weakly coupled  fermions \footnote{These include an excellent candidate for dark matter \cite{us}.}, a weak triplet of new gauge bosons, and a scalar triplet. The underlying strong dynamics is well hidden until much higher energies.

\section*{Acknowledgments}
This talk was based on work partially supported by the DOE under
contract DE-FGO3-96-ER40956, and  done in collaboration with Nima Arkani-Hamed,  Andy Cohen,  Emmanuel  Katz, Jaeyong Lee  and Devin Walker.

\end{document}